\def\year{2019}\relax
\documentclass[letterpaper]{article} 
\usepackage{aaai19}  
\usepackage{times}  
\usepackage{helvet}  
\usepackage{courier}  
\usepackage{url}  
\usepackage{graphicx}  
\usepackage{color}
\usepackage{todonotes}
\definecolor{notSureColor}{RGB}{147, 9, 229}
\colorlet{ColorforSina}{green!5!orange!95!}

\usepackage{etoolbox}
\makeatletter
\patchcmd{\maketitle}{\copyright@on}{}{}{}
\makeatother

\begin{document}
%

\title{Combating Fake News with Interpretable News Feed Algorithms}


\author{
  Sina Mohseni\\
    Texas A\&M University\\
    College Station, TX\\
    sina.mohseni@tamu.edu
  \And
  Eric D. Ragan\\
  University of Florida\\
  Gainesville, FL\\
  eragan@ufl.edu
}

\maketitle

\begin{abstract}
Nowadays, artificial intelligence algorithms are used for targeted and personalized content distribution in the large scale as part of the intense competition for attention in the digital media environment.
Unfortunately, targeted information dissemination may result in intellectual isolation and discrimination.
Further, as demonstrated in recent political events in the US and EU, malicious bots and social media users can create and propagate targeted ``fake news'' content in different forms for political gains.
From the other direction, fake news detection algorithms attempt to combat such problems by identifying misinformation and fraudulent user profiles.
This paper reviews common news feed algorithms as well as methods for fake news detection, and we discuss how news feed algorithms could be misused to promote falsified content, affect news diversity, or impact credibility.
We review how news feed algorithms and recommender engines can enable confirmation bias to isolate users to certain news sources and affecting the perception of reality.
As a potential solution for increasing user awareness of how content is selected or sorted, we argue for the use of interpretable and explainable news feed algorithms.
We discuss how improved user awareness and system transparency could mitigate unwanted outcomes of echo chambers and bubble filters in social media.
\end{abstract}
\section{Introduction}

The popularity of social media has resulted in large, continuously updating collections of user data.
Alongside the human users' activity in sharing information to socialize and express their opinions in a virtual social environment, machine learning algorithms actively process social media content for different purposes in very large scale.
Early examples include recommendation and personalization algorithms used in online e-commerce websites to make relevant product suggestions.
Similar approaches found their way into social media to assist users with finding preferred content and viewing more relevant advertisements.
With the growing amount of user data in social media, giant tech companies started transferring user preference data between different sectors and even third-party companies.

Meanwhile, the public as well as lawmakers have realized the importance of privacy and the influence of social media algorithms in serious matters.
Lawmakers such as the EU General Data Protection Regulation (GDPR) started legislation to protect user data transfer without consent.
The implications of personalized data tracking for the dissemination and consumption of news has caught the attention of many, especially given evidence of the influence of malicious social media accounts on the spread of fake news to bias users during the 2016 US election~\cite{bessi2016social}.
Recent reports show that social media outperforms television as the primary news source~\cite{allcott2017social}, and the targeted distribution of erroneous or misleading ``fake news'' may have resulted in large-scale manipulation and isolation of users' news feeds as part of the intense competition for attention in the digital media space~\cite{Social2018Inequalities}.

Although online information platforms are replacing the conventional news sources, personalized news feed algorithms are not immune to bias~\cite{bozdag2013bias} and can even cause intellectual isolation~\cite{pariser2011filter} over time.
Despite common beliefs that automated processes can reduce the chance of human judgment errors and biases, artificial intelligence systems are susceptible to accumulating bias from multiple sources.
Examples of sources of bias for machine learning systems include human bias in the creation of training data, imbalances in dataset distribution, or biases in the engineering and training processes.
Biased algorithms may cause unintentional discrimination that may result in a loss of opportunities (e.g., in employment, education, or housing opportunities), economic loss (e.g., in loan and credit application approval and differential prices for goods) and social stigmatization (e.g., loss of liberty and stereotype reinforcement) in large scale.
It is remarkable that despite all these limitations, lack of end-user awareness results in the blind trust of the user to AI systems.
For instance, a study~\cite{rader2015understanding} on users beliefs and judgments about Facebook's news feed algorithm shows a significant lack of awareness in users on how the system works and whether if the news feed algorithm is biased or not.

Transparency could be a solution in such circumstances by explaining why a certain recommendation or decision is made, but many advanced machine learning algorithms lack transparency and interpretability.
Lack of transparency in machine learning algorithms has been studied before the creation of deep neural networks (e.g.,~\cite{swartout1983xplain}), and research shows a trade-off between performance and transparency of the machine learning systems~\cite{eddy2006accuracy}.
With the recent advancement of deep learning algorithms, researchers are further emphasizing the necessity of interpretable machine learning systems~\cite{gunning2017explainable}.
Machine learning explanations could help end users understand how systems works and why certain algorithmic results are generated.
Recommendation algorithms were among the first machine learning algorithms to provide explanations to help users build trust.
Examples of explainable recommendation algorithms include movie and other product recommendations in e-commerce and entertainment platforms~\cite{Berkovsky:2017:RUT:3025171.3025209}.
While news feed algorithms are basically recommendation systems for news content, their function is similar to decision-making algorithms since the user has no chance (or very little) to choose between multiple choice of news.
Lack of explanation on how the content is selected for the user may result in unaware users who think they have access to all available information rather than only a small subset of news.
Explanation of the composition of news feeds and personalized search engines can help users understand how their news content is selected and if any bias is included in the results.

\begin{figure}[t!]
\centering
  \includegraphics[width=0.99\columnwidth]{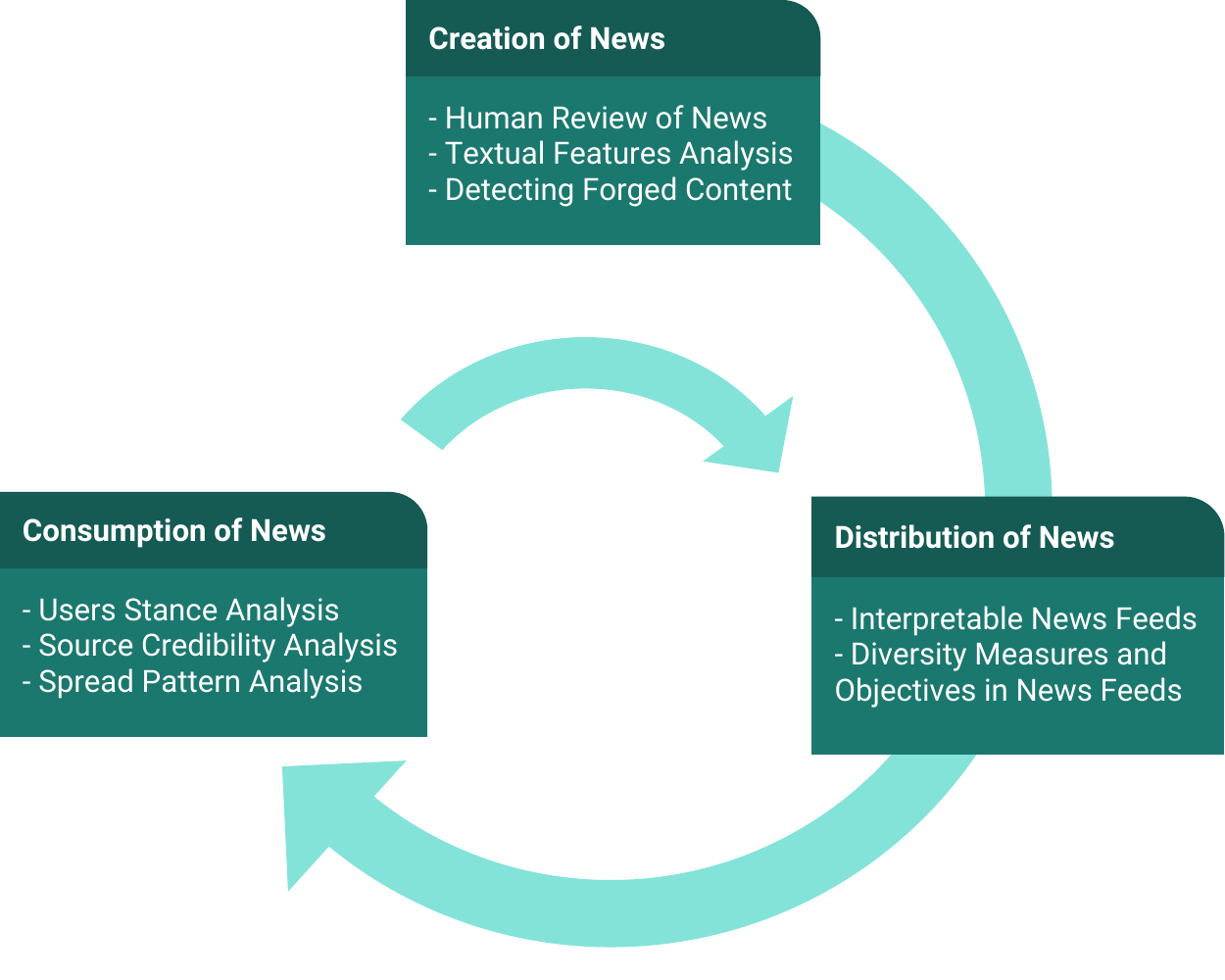}
  \caption{A summary of fake content detection methods at different stages of the news life in social media. 
  While natural language processing and social media data mining methods are popular in analyzing news at the consumption stage, there is a limited amount of research on accountable news recommendation algorithms to manage the propagation of fake content.
  The inner arrow shows how user's social data is used to update personalized news feed algorithms.
  }
  \label{fig:chain}
\end{figure}




In this paper, we first review how fake news and other falsified contents are being managed in online/social media.
We then categorize common methods for detecting the creation and circulation of falsified content in social media.
We also discuss how news feed algorithms may promote falsified content, impact news diversity, and hinder credibility.
We hypothesize that news feed algorithms are likely to align with user confirmation bias to isolate users to specific news source and perception of reality.
After all, these algorithms are designed to encourage users to click and increase interest and profit from advertisements. 
Finally, we also discuss interpretable news feed algorithms as a potential solution to increase user awareness and news diversity in social media.


\section{Analyzing Fake Content in Social Media} 

Fake content is being generated intentionally for the purpose of various political and financial benefits.
Although dishonest/biased coverage of political news and false commercial advertisement are among traditional human-made fake content, nowadays machine learning algorithms can also generate fake content (e.g., falsified videos\footnote{https://github.com/deepfakes/faceswap} and images, and text \cite{blom2015click}) with targeted user groups and at a very large scale in social media.
Trolls and malicious users also actively spread fake content to disrupt online communities.
Shu et al.~\shortcite{shu2017fake} provides a comprehensive review of data mining methods, recognition metrics, and training data sets used in fake news detection research.
However, their work acknowledges a missing piece in current research on detecting and managing fake news content in social media. 
We point to a research gap in studying effects of news recommendation algorithms on spread of fake content, crediting unreliable sources, and polarizing users feed (see Figure~\ref{fig:chain}). 
In this section, we review human review and data mining approaches to recognize fake content and open a discussion on the impact of news recommenders and distribution algorithms.

\subsection{Human Review}
Human review is the first category and manual process of analyzing content to determine its accuracy.
Fact checking is a knowledge-based approach to judging the veracity of news piece with external references~\cite{vlachos2014fact}.
Experts review the truthfulness of the news by evidence and determine whether claims are accurate or false (partially or entirely).
Fact checking is not limited to the correctness of textual content---images and videos may be evaluated or used as evidence as well.
As an example of human fact checking, Politifact\footnote{http://www.politifact.com/} is a non-profit fact checking organization that gathers experts to review daily news by investigating related reports and interviews in order to classify the relative accuracy of the news using a six-point rating scale. 
However, expert review fact checking methods are time-consuming, expensive, and not scalable for stopping the spread of fake content in social media. 

To further increase the scalability of the expert review method,  visual analytics tools can help experts in detecting social streams and track their evolution over time. 
For instance, LeadLine~\cite{dou2012leadline} is an interactive visual analytics system designed to help identify meaningful events in the news and enable experts to visually review temporal changes of these events.
Other tools also visualize summaries of social streams on topical and geographical aspects~\cite{wu2018streamexplorer}.
Crowdsourcing techniques have also been employed to reduce fact checking costs by collecting the wisdom of the crowd.
To solicit help from active users, some social media platforms are also including options for users to report suspicious posts and accounts to be reviewed later as a way of leveraging of the power of crowd.

\subsection{Data Mining}

Falsified information and fake news may also be detected by analyzing their content using various types of data mining and machine learning techniques.
One approach, for instance, is to use linguistic features to analyze writing styles to detect possible false content~\cite{afroz2012detecting}.
For example, recognizing deception-oriented~\cite{rubin2015truth} and hyper-partisan content~\cite{potthast2017stylometric} can be used as a basis for detecting intentionally falsified information. Researchers use existing falsified content on social media (e.g., from twitter~\footnote{$https://about.twitter.com/en_us/values/elections-integrity.html$}) and fact checking organizations (e.g., PolitiFact~\cite{wang2017liar}) as training data for such methods.
Another approach is to use clickbait detection algorithms~\cite{potthast2016clickbait} to analyze inconsistency between headlines and content of the news for possible fake news detection.
Image and video verification methods including deep learning algorithms (e.g.,~\cite{julliand2015image,afchar2018mesonet}) have also been explored to recognize falsified visual content.

In addition to analyzing the news by their content, other social media information such as source credibility~\cite{castillo2013predicting}, users'
stance~\cite{jin2016news}, and news temporal spreading pattern
~\cite{kwon2013prominent} have been used to assess the veracity of the news.
Such social features can be applied to user groups to evaluate the credibility of specific news pieces by considering the stance of a group of users for the news topics~\cite{tacchini2017some}.
Similarly, rumor detection methods aim to detect and track posts discussing a specific topic~\cite{zubiaga2018detection}.
Although most mentioned data mining methods do not perform direct fake news detection, these methods leverage social and textual feature to identify suspicious news pieces for human review (crowdsourced or expert review).
To increase fake news detection accuracy, training models for analyzing multi-source and multi-modal data is suggested as another approach.
For example, Shu et al.~\shortcite{shu2017exploiting} explored the correlation between news publisher bias, user stance, and user engagement together in their Tri-Relationship fake news detection framework.
In the following work, Shu et al.~\shortcite{shu2018fakenewsnet} proposed a training dataset to include news content and social context along with dynamic information of news and user accounts.

Despite the research on detecting the falsified content by its textual and social features, machine learning study on the importance of news recommendation algorithms in distribution of false content is very limited.  
The importance of news feed algorithms is their critical role in providing news content and their vulnerability for being misused and causing discrimination.
For the rest of this paper, we discuss the accountability of news feed algorithms for propagating fake content by creating echo chambers and filter bubbles in social media and news search engines.
We suggest interpretable news feed algorithm will solve these problems by increasing user awareness, holding users accountable for their content, and exposing users to opposite political views.

\section{Limitations of News Feed Algorithms}

Although recommendation algorithms were primarily designed as adaptive systems for recommending, sorting and filtering search result entertainment and e-commerce platforms, they found their way into the news recommendation and search engines to provide news based on users interest.
Nowadays, personalized search engines and news recommendations on social media and provide content to users based on users' profiles, interests, social media friends, and other past click behavior.
However, research shows the personalization feature may create adverse situations such as filter bubble and echo chamber effect by promoting contents which are in favor of the user's existing mindset and eliminating conflicting viewpoints and their sources.
Further, lack of transparency in these algorithms may cause discrimination (e.g., by providing biased recommendations) and even open the door to targeted distribution of malicious paid contents to bypass regulations.
In this section, we review the downsides of news feed and personalized news search algorithms that may propagate and accredit of fake news.


\subsection{Echo Chambers in Social Media}

Echo chamber in social media describes a situation where homogeneous views are reinforced by communication inside closed groups.
Research shows the personalization feature can create 
the ``echo chamber'' effect similar to echo chambers in social media~\cite{quattrociocchi2016echo} by promoting contents which are in favor of the user's existing mindset.
Echo chambers have been previously studied in relation to creating polarized opinions and shaping a false sense of credibility for users who frequent news sources through social media~\cite{zajonc2001mere,paul2016russian}.
This false sense of credibility holds users in a vulnerable position of accepting biased and fake news content.
In similar circumstances, although news recommendation algorithms are meant to provide content related to users interest, these machine learning algorithms can isolate a user's news feed with a certain perception of reality and trigger user confirmation bias to over-trust partisan sources. 

Investigating such phenomena, a recent study by Geschke et al.~\shortcite{geschke2018triple} presented a simulation of different information filtering scenarios that may contribute to social fragmentation of users into distinct echo chambers.
Their observation of agent-based modeling found that social and technological filters can boost social polarization and lessen the interconnections of social media echo chambers.
This evidence clarifies the potentially negative impact of news recommendation algorithms in creating more distinct echo chambers which calls for further investigation of solutions.
Others study methods to overcome echo chambers in social media.
For instance, Lex et al.~\shortcite{lex2018mitigating} presented a content-based news recommendation that can increase exposure of the opposite view to users in order to mitigate the echo chamber effect in social media.
They computed and compared the diversity of the news on Twitter by their hashtags in different modes of a news recommender to balance users' news feeds and prevent the creation of echo chambers.
In another work, Hou et al.~\shortcite{hou2018balancing} demonstrated methods to balance popularity bias in network-based recommendation systems and to significantly improve the system's diversity and accuracy.

\subsection{Filter Bubbles in Search Engines} 

\textit{Filter bubble}~\cite{pariser2011filter} is another term to describe negative effects of personalized search engines and news feeds.
Filter bubbles represent a state of intellectual isolation where users are only exposed to a certain perspective of information brought to them by personalized search engines and news feed algorithms.
The lack of news diversity and exposure to conflicting viewpoints in the long term creates a filter bubble for individuals and social media groups and increase the chance of accepting misleading information and accrediting unreliable sources. 

In their study of targeted news in Facebook, Bakshy et al.~\shortcite{bakshy2015exposure} discuss inconclusive conclusions about the role of news feeds in creating biased social media environments.
However, other work indicates negative effects of personalization and explores ways to mitigate these effect.
For example, Nguyen et al.~\cite{nguyen2014exploring} measure content diversity at individual user level in a longitudinal study on collaborative filtering-based recommender system.
They contribute a new metric to measure content diversity and their study on a movie recommendation system indicates that recommender algorithms expose users a narrower set of items over time.
In another work, Haim et al.~\shortcite{haim2018burst} conducted an exploratory analysis of personalized news search engine effect on news diversity.
They studied both explicit (i.e., user-defined filters) and implicit personalization (i.e., by giving algorithms the chance to observe a given agent during a week-long period), and their results showed a general bias toward over-presenting certain news outlets and under-presenting other sources. 

Another approach to encounter unwanted negative effects of these algorithms is to define new measures and standards for sensitive data and products recommendation systems.
For better measuring the quality of recommender systems, Ekstrand et al.~\shortcite{valcarce2018robustness} and Valcarce et al.~\shortcite{ekstrand2018all} proposed new evaluation measures to account for other considerations like users' popularity bias or content diversity.
The importance of new measures for news recommendation algorithms comes from considering the diversity of content across different user groups. 

\subsection{Biased Algorithms and Paid Content}

Algorithmic bias is another issue with news and content recommendation systems.
Example of biased news recommendation system is to eliminate content from a certain geographic area and not by a thorough analysis of the content itself.
The importance of algorithmic bias is more sensible since the bias could enter the algorithms from different sources including bias from the training data, data cleaning and engineering, and bias in training.

Multiple works study bias and discrimination in algorithms via different techniques such as auditing~\cite{sandvig2014auditing} to investigate algorithms that are still working well.
Researchers also propose using diversity metrics~\cite{valcarce2018robustness,ekstrand2018all} and bias quantification methods~\cite{kulshrestha2017quantifying} as other ways to study bias and discrimination in recommendation algorithms.
For example, Kulshrestha et al.~\shortcite{kulshrestha2017quantifying} proposed a framework to quantify bias in ranked search results in political-related queries in Twitter.
Their framework can distinguish bias from news content and ranking algorithm, and they found evidence of significant effects of both input content and search algorithms in producing bias.

Further, paid advertisements can enter feeds for targeted purposes.
Paid content that is regularly in the form of commercial advertisements may also create opportunities for malicious accounts to spread false content to manipulate users for political gains.
In a sense, paid content enters users' news feeds from a back door along with related social posts and news to attract user attention and increase impact.
Recent examples of distributing paid content are seen in the influence of Cambridge Analytica, a political consulting firm, for political advertisements in social media.
They attracted heavy attention and criticism for their use of psychographic-profiling methods on Facebook user profiles in order to perform unique voter-targeting models during the United States presidential campaign and the Brexit campaign~\cite{persily20172016}.

Along with the aforementioned weaknesses of news recommendation algorithms, research has shown many users are not aware of the existence of news feed algorithms or do not know how such algorithms work.
Results from Rader and Gray's~\shortcite{rader2015understanding} study on how individuals make sense of the Facebook's news feed algorithm shows user understanding of ``how the news feed works'' is limited to their repeated experience with the system.
We argue that increasing user awareness by transparent news search engines and news feed algorithms not only will improve the users understanding of these algorithms, but also increases the diversity of news in social media.
Transparency and accountability in such large-scale employed algorithms is essential to detect possible algorithmic bias and user data misuse to distribute fake content and cause discrimination.
\begin{figure*}[t!]
\centering
  \includegraphics[width=1.7\columnwidth]{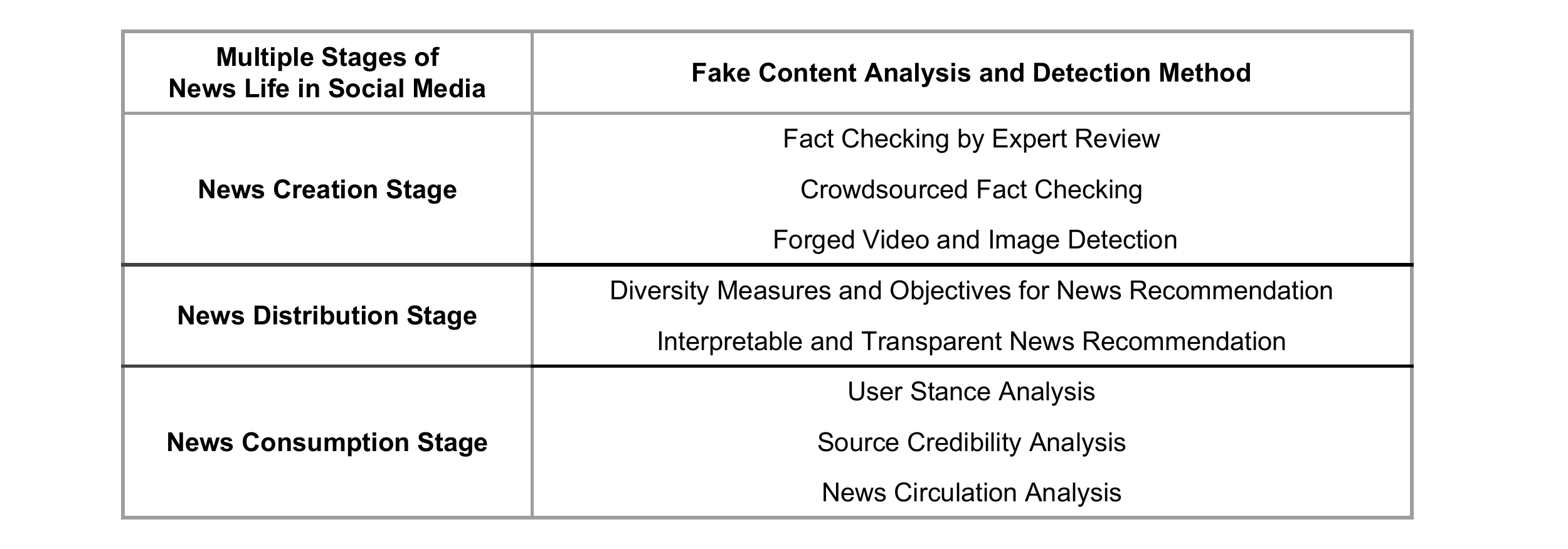}
  \caption{Different approaches to detect and manage falsified content and fake news in social media. While most data mining methods aim to identify false information that is already spread in social media, interpretable and transparent news feed algorithms can manage the propagation of falsified content before it spreads.
  }
  \label{fig:table}
\end{figure*}

\section{The Benefits of Interpretability}

In the previous sections, we discussed current methods to detect and manage fake content in social media.
We also discussed how news feed algorithms and personalized search engines may contribute to spreading fake content by not showing diverse news.
In this section, we propose interpretable and accountable news feed algorithms as a potential solution by increasing news diversity and improving user control over readers' content.

Interpretable machine learning algorithms generate explanations and describe the reasoning behind machine learning decisions and predictions.
Machine learning explanation enables the user to understand how the data is processed and supports user awareness to identify possible bias and systems malfunctions.
Interpretable machine learning research has many applications in algorithmic decision making such as fair and ethical computing and accountability in AI~\cite{doshi2017towards}. 
For example, to measure users' perception of justice in intelligent decision making, Binns et al.~\cite{Binns:2018:RHP:3173574.3173951} studied explanations in daily-life systems such as determining car insurance rates and loan application approvals.
Their results highlight the importance of machine learning explanations in users' comprehension and trust in algorithmic decision-making systems.
Recommendation algorithms have been among the earliest algorithms to include explanations to improve users trust and experience.
Explanation in recommender systems have been shown to increase recommendation uptake by improving system's transparency and user's trust.
In regard to recommendation systems in real-world applications, Berkovsky et al.~\shortcite{Berkovsky:2017:RUT:3025171.3025209} studied user trust with various recommendation interfaces and content selection strategies.
They investigated user reliance on a movie recommender system in various conditions of recommendation interfaces.

However, machine learning explanations in product recommender systems are limited to providing product information, highlighting users need by their search history, and other domain-specific features.
In contrast, news recommendation algorithms in social media often process more complicated and sensitive data. 
Also, news recommendations generate a user's reading list and do not give options for the user to choose among content. 
In these circumstances, lack of explanation on how the content is selected for the user may bias the user toward what the algorithm is providing in the news feed and not the entire reality.


Interpretable news feed algorithms can help users understand how the information selection process is done and allow users to review a system's objective.
Explaining the user preference model and other parameters (e.g., the priority of paid content) that contribute to the news selection will help users to have a more accurate mental model of these algorithms and build appropriate trust to the system.
In a study on the effects of news feed explanations on user awareness, Rader et al.~\shortcite{rader2018explanations} showed complementary explanations for the news feed algorithm helped users to determine if the news feed algorithm was biased or if they had control over the content.
In their study, they measured users' awareness, correctness, and accountability to evaluate algorithmic transparency.
Although they used a general explanation for how Facebook's news feed works, their results showed an increase in user awareness with explanations.

We argue that further detailed explanations of a news recommendation algorithm (e.g., in form of visualization of model parameters) enables the user to detect biased models and suspect odd outputs.
Such an explanation would tell the user what kind of \textit{soft filters} are applied on their news feed to select the user content.
Related to this, Burbach et al.~\shortcite{burbach2018user} studied user preferences of recommendation filters through an online survey.
Their survey shows users have different patterns of recommendation method preference for different product categories. 
A significant finding was that in more sensitive categories (e.g., news recommendation) users preferred content-based recommendations and rejected social-based methods that require personal data to process.
In other work, Rader and Gray's~\shortcite{rader2015understanding} study showed that the mismatch between the objective of those who design recommendation algorithm and the object of the user may lead to reduced user interests in social media platforms. 
Therefore, empowering the user to adjust and prioritize the influence of recommendation could also have positive outcomes for long-term acceptance of social media.
To achieve accountable news feed algorithms, it may be sufficient to allow users to eliminate unrelated and unintended content by reviewing and adjusting the algorithm's parameters, but designing usable explanation interfaces to allow such functionality may be challenging.

Table~\ref{fig:table} shows a list of potential solutions to manage fake news content in social media at different stages of news life.
Detecting fake content with human review can be done through expert review and crowdsourcing techniques at the early stages.
Also, computer vision techniques help in detecting forged images and videos known as DeepFakes~\cite{afchar2018mesonet}.
New methods like provenance analysis have also been utilized for content validation via generating provenance graph of images as the same content is shared and modified over time~\cite{bharati2018beyond}.
The next stage of fake news life in social media is the distribution of content via search engines and news feed algorithms.  
Others also examined the use of spacial news feed in virtual reality environment as opposed to linear feeds to reduce the creation of filter bubbles~\cite{linder2018pop}.
Multiple sources of evidence show personalized news feed algorithms and targeted advertisement of political content can have drastic effects on news diversity and the creation of echo chambers in social media.
The final stage of analyzing fake news is to process social media users' stance, analyze news propagation patterns, and estimate news source credibility to find possible false content in social media.
Although research on social media data mining shows that combinational methods help to reduce the cost of computation and increase fake news detection accuracy~\cite{shu2017exploiting}, using social media data means waiting until the fake content is already exposed to the users.

Looking at the current research in fake news and current advancements in interpretable machine learning algorithms, we suggest that adding a layer of explanation to news recommendation algorithms would serve as a major leap toward accountable news streaming algorithms and eliminate propagation of fake content by increasing user awareness and transparency in social media news feed and news search engines.

\section{Conclusion}
We reviewed fake news in social media over three main stages and discussed current methods and tools for analyzing and detecting potentially false content.
We then focused on limitations and risks (e.g., misuse of user data) of using personalized content selection algorithms in social media.
We argue that the large-scale targeted propagation of content for political gain using personalized news feed may cause polarized social media and spreading of fake content.
After reviewing current research on fake news detection and personalized news feed algorithms, we suggest interpretable news feed algorithms as a potential solution to increase user awareness and hold users accountable for their own content rather than algorithms with undisclosed/opaque objectives.


\bibliographystyle{aaai}
\bibliography{Bibliography}

\end{document}